\documentclass{PoS}

\title{Monopole condensation in two-flavour Adjoint QCD}

\ShortTitle{Monopole condensation in $N_f=2$ aQCD}

\author{\speaker{Giuseppe Lacagnina}\\
        Dipartimento di Fisica and INFN, Pisa\\
        E-mail: \email{lacagnina@df.unipi.it}}

\author{Guido Cossu\\ Scuola Normale Superiore and INFN, Pisa\\ E-mail: \email {g.cossu@sns.it}}
\author{Massimo D'Elia\\ Dipartimento di Fisica and INFN, Genova\\ E-mail: \email{Massimo.Delia@ge.infn.it} }
\author{Adriano Di Giacomo, Claudio Pica\\ Dipartimento di Fisica and INFN,
  Pisa\\ E-mail: \email{digiaco@df.unipi.it}, \email{pica@df.unipi.it}}

\abstract{Two distinct phase transitions occur at different temperatures in
  QCD with adjoint fermions (aQCD): deconfinement and chiral symmetry
  restoration. In this model, quarks do no explicitely break the center Z(3)
  symmetry and therefore the Polyakov loop is a good order parameter for the
  deconfinement transition. We study monopole condensation by inspecting the
  expectation value of an operator which creates a monopole. Such a quantity
  is expected to be an order parameter for the deconfinement transition as in
  the case of fundamental fermions.}

\FullConference{XXIV International Symposium on Lattice Field Theory\\
                 July 23-28 2006\\
                 Tucson Arizona, US}

\begin{document}

\section{Introduction}

As lattice simulations seem to show, chiral symmetry restoration and
deconfinement occur for QCD at the same temperature, making it extremely hard
to disentangle them. On the other hand, QCD with quarks in the adjoint
representation of $SU(3)$ (aQCD) is a model in which two transitions take
place at different temperatures \cite{karsch}. In aQCD, the $Z(3)$ center
symmetry is not broken by adjoint fermions and therefore the Polyakov loop is
an order parameter even at finite quark mass as in the pure gauge case.\\

The authors of ref.\cite{karsch} find on the lattice two distinct phase
transitions; they observe a first order deconfinement transition and a
continuous chiral transition. They also check that the observables which are
sensitive to deconfinement are not significantly affected by the chiral transition.\\

Another way to study confinement consists in looking for magnetic charge
condensation, which signals dual superconductivity of the vacuum
\cite{pisa1,pisa2, pisa3, pisa4}. In this case, one constructs an operator
which carries magnetic charge and looks at its vacuum expectation value, which
is expected to be different from zero in the confined phase and strictly zero
in the deconfined phase. Our goal is to study monopole condensation in aQCD
in order to check that it is unaffected by the chiral transition.\\

In this report we present the results of a lattice simulation of aQCD in which
we started our investigation; it has to be considered as work in progress. The
qualitative behaviour of the monopole order parameter was found to be
consistent with the expectations; however, for time limitations, we could not
simulate quark masses that were light enough to study its chiral properties.
We are currently continuing our investigation in this direction. We also need
to understand the scaling properties of the monopole condensation parameter.

\subsection{aQCD}

Quarks in the adjoint representation of QCD have $8$ colour degrees of freedom
and can be described with $3\times 3$ hermitian traceless matrices
\begin{equation}
Q(x) = Q^a(x)\lambda_a
\end{equation}
using Gell-Mann's $\lambda$ matrices. In order to write the fermionic part of
action for this model, the $8-$dimensional representation of the gauge links
(which is real) must be used:
\begin{equation}
U^{ab}_{(8)} = \frac{1}{2}{\rm Tr}(\lambda^aU_{(3)}\lambda^bU_{(3)}^{\dagger})
\end{equation}
The full action is therefore given by
\begin{equation}
S = S_G[U_{(3)}] + \sum_{x,y} {\bar Q}(x)M(U_{(8)})_{x,y}Q(y)
\end{equation}
where $S_G$ is the usual $SU(3)$ gauge action and $M$ is the fermionic
matrix. The Polyakov loop is defined as in the $SU(3)$ Yang Mills case
\begin{equation}
L_{3} \equiv \langle \frac{1}{3L_s^3}|\sum_{{\vec x}}{\rm
  Tr}\prod_{x_0=1}^{L_t}U_0^{(3)}(x_0,{\vec x})|\rangle
\end{equation}
This quantity is an order parameter for the spontaneous breaking of the center
symmetry and is related to the free energy of isolated static fundamental
charge.
\subsection{Monopole condensation}

A magnetically charged operator is constructed in such a way that it adds a
monopole field to a given configuration. Its expectation value is given by
\begin{equation}
\langle \mu \rangle = \frac{\tilde{Z}}{Z}
\end{equation}
where $\tilde{Z}$ is the partition function for the action in presence of a
monopole. To better cope with fluctuations, one instead calculates the quantity
\begin{equation}
\rho = \frac{d}{d\beta} \ln \langle \mu \rangle = \langle S\rangle_S - \langle\tilde{S}\rangle_{\tilde{S}}
\end{equation}
in which $\langle\tilde{S}\rangle_{\tilde{S}}$ is the average of the action
with a monopole insertion weighted with the modified action itself. Therefore,
two simulations have to be run for each value of $\beta$. The parameter
$\langle \mu\rangle$ should be different from zero in the confined phase where
magnetic charges condense and drop to zero at deconfinement where magnetic
symmetry is restored. This drop corresponds to a negative peak of $\rho$. In
the vicinity of the critical temperature, $\beta\simeq\beta_c$, $\rho$ is
expected to scale as
\begin{equation}
\rho L^{-1/\nu}=f\left( L^{1/\nu}(\beta_c-\beta)\right)
\end{equation}
The critical exponent $\nu$ should be equal to $1/3$ for a first order
transition.

\section{Simulation details}

We have simulated two flavours of adjoint staggered fermions using the RHMC
algorithm \cite{rhmc}.  Trajectories had a length of $N_{MD}\delta t=1$, and
an integration step of $\delta t=0.02$. Acceptance was above $90\%$.
Inversions of the fermionic matrix were performed with the Conjugate Gradient
algorithm.

We have run simulations with two different lattice sizes, i.e. $L_s^3\times
L_t=12^3\times 4, 16^3\times 4$. We have calculated the chiral condensate, the
Polyakov loop and the average plaquette for $12$ values of $\beta$ in the
range $(3.0,7.0)$. We computed the average plaquette and the $\rho$ parameter.
For time constraints, we were limited to runs at a bare quark mass of
$am=0.04$. In order to simulate an action with a monopole contribution, $C^*$
boundary conditions had to be implemented \cite{cstar}.

Our code has been run on the ApeMille machine in Pisa and the ApeNEXT facility
in Rome.

\section{Results}

\begin{figure}
\includegraphics[width=0.95\textwidth]{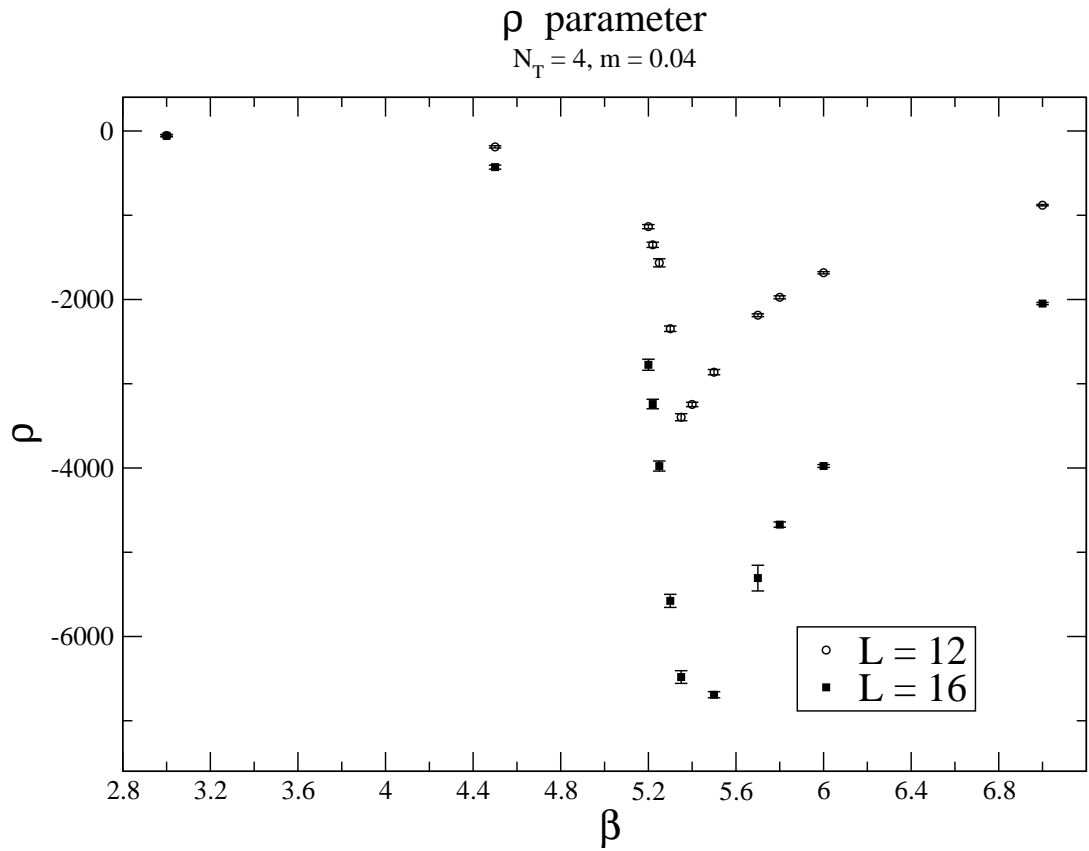}
\caption{The $\rho$ parameter, with $am=0.04$, $L_t=4$, for two different
  spatial volumes}
\label{plot_rho}
\end{figure}

\begin{figure}
\includegraphics[width=0.95\textwidth]{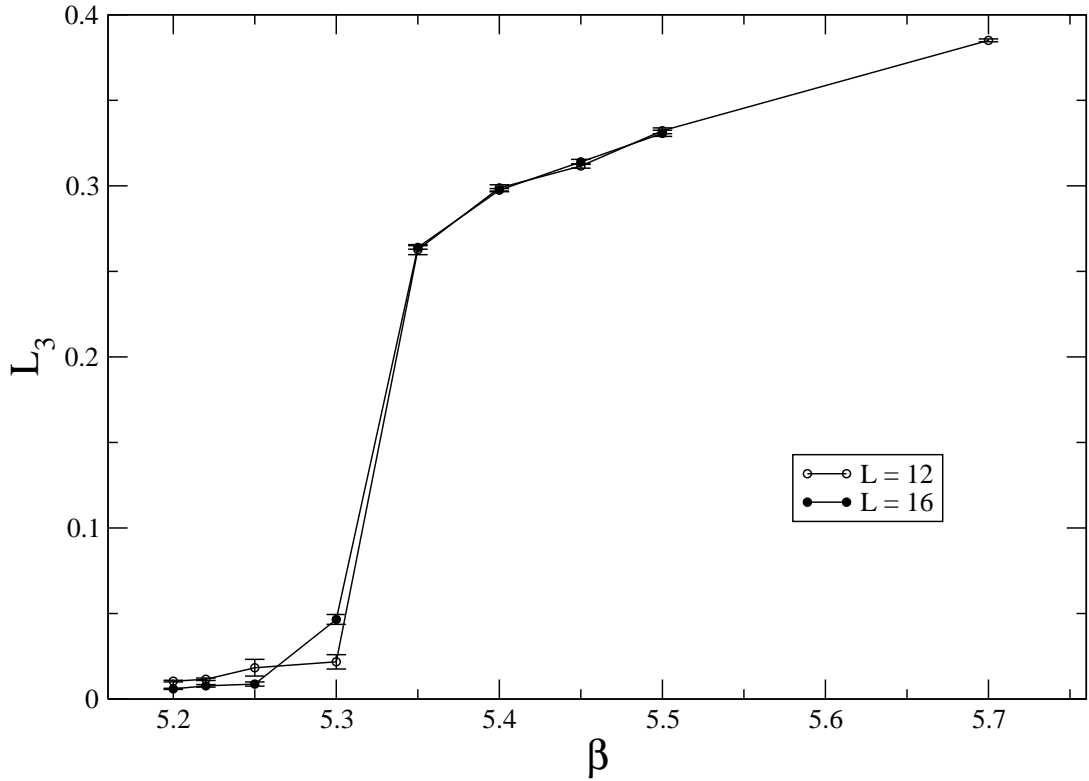}
\caption{The Polyakov loop, with $am=0.04$, $L_t=4$, for two different
  spatial volumes. Lines are left to guide the eye.}
\label{polyakov}
\end{figure}

For each value of $\beta$, we have calculated the Polyakov loop and the $\rho$
parameter on both volumes, $12^3\times 4, 16^3\times 4$. For the Polyakov
loop, we reproduced the results of \cite{karsch} at the same value of the bare
quark mass, $am=0.04$, see Figure (\ref{polyakov}). Figure (\ref{plot_rho})
shows the results for the $\rho$ parameter. The position of the negative peak
shifts slightly with the volume as one would expect for a physical transition.
This estimate of the pseudocritical coupling is consistent with the position
of the jump in the Polyakov loop. Furthermore, as $\beta\rightarrow 0$, one
expects $\rho$ to become volume independent, as $\mu$ should tend to a nonzero
constant. Our results are compatible with this expectation. We are currently
working on the analysis of the scaling properties of $\rho$.  We could not
study the behaviour of $\rho$ at the chiral transition as the bare quark mass
turned out to be too large to see any visible effect in the chiral condensate
susceptibility.

\section{Conclusions}

We have studied monopole condensation in lattice QCD with $N_f=2$ staggered
fermions in the adjoint representation. We evaluated the expectation value of
a magnetically charged operator, which is expected to be an order parameter
for confinement. For this operator, we observed a qualitative agreement with
the expected behaviour. In particular, the data were consistent with the
theoretical predictions for the $\beta \rightarrow 0$ limit: $\rho$ becomes
volume independent. We are currently working on simulations with lighter quark
masses and larger spatial volumes, in order to study the behaviour of the
$\rho$ parameter at the chiral transition and its scaling properties.

\end{document}